\documentclass[12pt]{iopart}
\usepackage{amsmath,bm,amssymb}
\usepackage{mathrsfs}
\usepackage{amsfonts}
\usepackage{graphicx}
\usepackage{color}
\usepackage{subfigure}
\usepackage{dcolumn}

\begin{document}
\title{Electronic structure of atomic manganese chains supported on Cu$_2$N / Cu (100)}
\author{Deung-Jang Choi\footnote{Corresponding author: d.choi@nanogune.eu}}
\address{CIC nanoGUNE , Tolosa Hiribidea 78, Donostia-San Sebasti\'an 20018, Spain}
\author{Roberto Robles}
\address{
Catalan Institute of Nanoscience and Nanotechnology (ICN2), CSIC}
\address{The Barcelona Institute of Science and Technology, Campus UAB, Bellaterra, 08193 Barcelona, Spain}
\author{Jean-Pierre Gauyacq}
\address{Institut des Sciences Mol\'eculaires d'Orsay (ISMO), CNRS, Univ.
 Paris-Sud, Universit\'e Paris-Saclay, B\^at. 351, 91405 Orsay CEDEX,
France}
\author{Carmen Rubio Verd\'u}
\address{CIC nanoGUNE , Tolosa Hiribidea 78, Donostia-San Sebasti\'an 20018, Spain}
\author{Nicol\'as Lorente}
\address{Centro de F{\'{\i}}sica de Materiales
CFM/MPC (CSIC-UPV/EHU), Paseo Manuel de Lardizabal 5, 20018 Donostia-San Sebasti\'an, Spain}
\address{Donostia International Physics Center (DIPC), Paseo Manuel de Lardizabal 4, 20018 Donostia-San Sebasti\'an, Spain}
\author{Jos\'e Ignacio Pascual}
\address{CIC nanoGUNE , Tolosa Hiribidea 78, Donostia-San Sebasti\'an 20018, Spain}
\address{Ikerbasque Basque Foundation for Science, 48013 Bilbao, Spain}

\begin{abstract}
Scanning tunnelling microscopy and density functional theory studies of
manganese chains adsorbed on  Cu$_2$N/Cu (100) reveal an unsuspected
electronic edge state at $\sim 1$ eV above the Fermi energy. This
Tamm-like state is strongly localised to the last Mn atom of the
chain and fully spin polarised. However, no equivalence is found
for occupied states, and the electronic structure at $\sim -1$ eV
is mainly spin unpolarised due to the extended $p$-states of the N
atoms that mediate the coupling between the Mn atoms in the chain.
Odd-numbered Mn chains present an exponentially
decreasing direct coupling with distance between the two edges,
leading to a vanishing bonding/anti-bonding splitting of states while even-numbered
Mn chains present perfect decoupling of both edges due to the the
antiferromagnetic ordering of Mn chains.  
\end{abstract}

\date{\today}

\maketitle

\section{Introduction}

Magnetic nanochains are experiencing a lot of interest due to
their quasi 1-D character that confers them with extraordinary
properties~\cite{wang_2011}.
Atomic magnetic nanochains are the best examples of what magnetic
nanodevices can achieve and how they can be instrumental for
spintronics~\cite{brune_2006}. 
These chains are assembled using the
atom manipulation capabilities of the scanning tunnelling
microscope (STM). Magnetic atoms have been positioned one by one
at different distances and with different arrangements on a variety of
substrates~\cite{hirjibehedin_2006,khajetoorians_2012,loth_2012,holzberger_2013,bryant_2013,yan_2014}.
The STM has permitted to characterize the chains by their spin signature
using spin-polarised tips~\cite{wiesendanger} and by their inelastic
electron tunnelling spectra (IETS)~\cite{heinrich_2004,PSS}. These
measurements give unprecedented insight into the atomic mechanisms
leading to magnetic ordering in nanostructures that can be compared with
state-of-the-art theoretical results.

Theoretical works are generally based on density functional theory
(DFT) studies. These works evaluate the actual atomic arrangements
of the atoms on the surface, the local and global magnetic moments,
as well as the magnetic anisotropy energies, the exchange couplings
among the chain constituents and the possibility of canting due to
the Dzyaloshinskii-Moriya interaction.  For the case of Mn chains on
Ni (100), Lounis and co-workers~\cite{lounis_2008} showed that the
competition of the different exchange couplings in the system led
to an even/odd effect with the number of Mn atoms; even-numbered
chains presenting a non-collinear arrangement of their spins and
chains with an odd number of atoms a collinear antiferromagnetic
ordering. Rudenko and collaborators~\cite{rudenko_2009} performed
thorough calculations of Mn chains on a Cu$_2$N/Cu (100) substrate.
They reproduced the exchange couplings between atoms that lead to
magnetic excitation spectra in good agreement with the experimental
ones~\cite{hirjibehedin_2006}. Furthermore, they included spin-orbit
interactions with different methods to study the non-collinearity of
the magnetic-moment distributions,  and they obtained that the Mn
atoms had an out-of-plane easy axis, and that the canting of spins
due to anisotropic exchange interactions was very small. Another
complete study of Mn and Co chains on Cu$_2$N/Cu (100) was performed
by Lin and Jones~\cite{lin_2011} where they extended their previous
results~\cite{hirjibehedin_2007} and confirmed that the atoms maintain
their nominal spins on the surface, $S$=5/2 for Mn.  Nicklas and
co-workers~\cite{nicklas_2011} studied Fe chains on Cu$_2$N/Cu (100)
showing that as for Mn,~\cite{rudenko_2009} N-mediated superexchange leads
to antiferromagnetic coupling of the Fe atoms, in good agreement with
later experimental measurements~\cite{loth_2012}.  The interpretation
of these experiments has shown the importance of correlation and
entanglement in these antiferromagnetic chains~\cite{gauyacq_2013}.
Urdaniz {\em et al.}~\cite{urdaniz_2012} performed a thorough study
of Cr, Fe, Mn and Co chains on Cu$_2$N/Cu (100) using DFT calculations
showing that the adsorption site determines to a great extent the type
of magnetic coupling of the chain.  This is presently used to generate
atomic chains with different coupling schemes~\cite{spinelli_2015}.
DFT calculations show how important it is to take into account the
actual geometries of the chain, because this can completely change their
electronic and magnetic behaviour~\cite{Cochains,Tao2015}.

All these works focus in the low-energy structure tunnelling conductance
spectra that has a direct link to the magnetic properties of the crafted
nano-objects.  Surprisingly, no work has been studied on the larger energy
scale that actually has influences on the magnetic properties of these
systems.  In the present work, we report on the electronic structure with
a special attention to states originating in orbitals more extended than
pure $d$-electrons.  We show that there are long-lived edge states that
maintain strict localisation. These edge states are Tamm states due to the
unsaturated bond of the edge Mn atom caused by the tilting of the last
N--Mn bond together with the different nature of the last N atom. This
last N atom presents a different environment (lack of Mn atom on one side,
and a closer N-substrate distance) breaking the symmetry of the chain.
The case of even-numbered Mn chains is particularly relevant for the
link between the edge state and the particular magnetic properties of
the chains. These edge states are at $\sim 1$ eV above the Fermi level. A
broader resonance is also found for occupied states at about $\sim -1$ eV.
However, there are no specific magnetic features associated with this
state and it is rather a state originating in the covalent bonding of
the Mn atoms with the N atoms glueing the Mn chain together.
The magnetic structure of the Mn chains are due to the spin
polarization of the $d$-electrons, much lower in energy than the  $\sim -1$ eV
structure of the N-Mn bonds.

\section{Experimental method}

Experiments were performed in an ultrahigh-vacuum low-temperature STM at
a base temperature of 1.15 K. The differential conductance was directly
measured using lock-in detection with a 2-mV rms modulation at 938 Hz
of the sample bias V. 

The Cu(100) surface was cleaned by Ar sputtering and then annealed up
to 650 K. After having big terraces of the Cu(100) crystal, a monolayer
of Cu$_2$N was formed as a decoupling layer by N irradiation. Single
 Mn atoms were deposited onto the cold surface.  By capturing the Mn atom with the
tip and dropping it onto the substrate via bias pulses, the single atoms
were arranged into closed-packed Mn chains along the [010]-direction
of the Cu$_2$N surface. This leads to mono-atomic chains of Mn atoms
ontop of Cu atoms that are aligned along a nitrogen row, identical to
the structures reported in Ref.~\cite{hirjibehedin_2006}.

\section{Theoretical method}
\label{theory}

{\it Ab initio} calculations were performed within the
density-functional theory (DFT) framework as implemented in the
VASP code~\cite{kresse_efficiency_1996}. We have expanded the
wave functions using a plane-wave basis set with a cutoff energy of
300~eV. Core electrons were treated within the projector augmented wave
method~\cite{bloechl_projector_1994, kresse_ultrasoft_1999}. The PBE
form of the generalized gradient approximation was used as exchange
and correlation functional~\cite{perdew_generalized_1996}.  To model
the surface we have used a slab geometry with four Cu layers plus the
Cu$_2$N layer.  We have used an optimized theoretically lattice constant
for Cu of 3.65~\AA.

Following the above experimental procedure, the transition-metal atoms
are positioned on Cu atoms, forming a chain in the [010] direction. We
have used a unit cell that increases its size along this direction with
the number of atoms of the chain as [3 $\times$ ($n$+3)], where $n$
is the number of Mn atoms. In this way we keep the distance between
chain images constant for all sizes, being of 3 lattice constants in
the unrelaxed configuration.  The bottom Cu layer was kept fixed and
the remaining atoms were allowed to relax until forces were smaller than
0.01~eV/\AA. The $k$-point sample was varied accordingly to the unit cell,
and tests were performed to assure its convergence.

In order to account for the atomic magnetic moments of the Mn 
atoms on the surface, the GGA+U method of Dudarev \textit{et al}~\cite{dudarev_1998} was
employed, with a $U_{\texttt{eff}}=U-J$ of 4~eV.
The chosen values correspond to roughly substracting $J\approx 1$ to
$U=4.9$ eV as computed by Lin and Jones~\cite{lin_2011} for Mn ontop a Cu atom. 

\section{Results}
%Experimental results
\subsection{Scanning tunneling spectroscopy of electronic states}

\begin{figure}[ht] \centering
\hspace{2cm}\hfill\includegraphics[width=0.8\columnwidth]{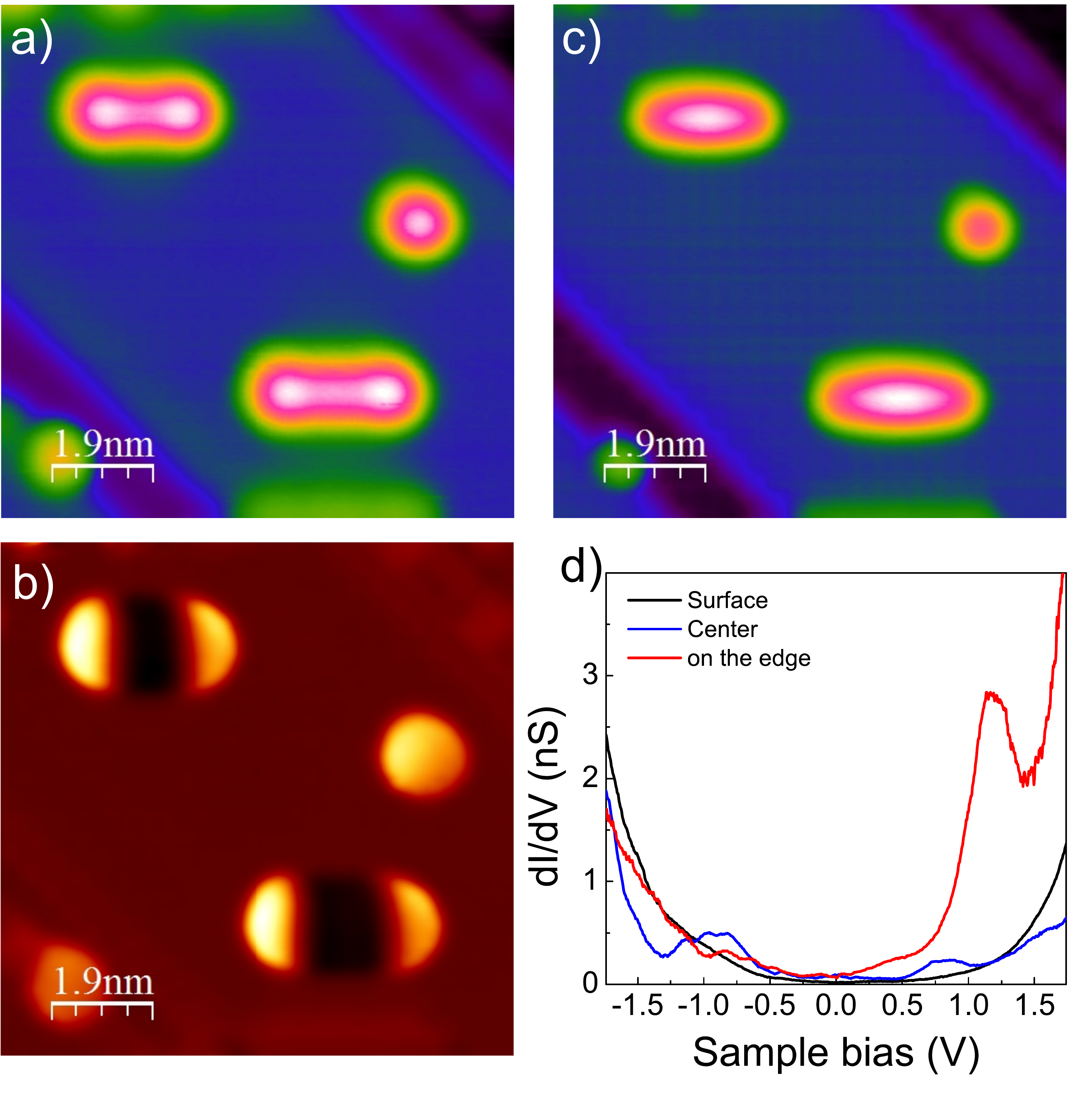} 
\caption{\label{stm}
{ $(a)$ Constant current image taken at 2~V showing a Mn$_5$ chain (upper
left corner) and a Mn$_6$ chain (bottom centre) (V$_s$ = 2~V, I$_t$ = 100~pA). $(b)$ Conductance map
at 1~V over the same area (V$_s$ = 1~V, I$_t$ = 1~nA, lock-in modulation 10mV rms at 938.6Hz). The prevailance of the edges of the chains
is clearly seen as maxima in the conductance.  $(c)$ The constant
current image at -1~V is rather featureless over all the chains (V$_s$ = -1~V, I$_t$ = 100~pA).
$(d)$ The tunnelling conductance as a function of applied bias over
three different spots on the surface (feed back opened at V$_s$ = -2~V, I$_t$ = 1.5~nA). The Conductance on the edge of
the Mn$_6$ chain clearly displays a maximum at $\sim$ 1 V and another
maximum at $\sim$ -1 V. At the centre of the chain the maximum at $\sim$
1 V is not present but the maximum at $\sim$ -1 V is slightly displaced.
For comparison the conductance of the clean Cu$_2$N/Cu (100) surface
is shown. } } 
\end{figure}

Constant current STM images obtained for sample biases above 1V show that the Mn chains develop
a ``dumbell'' shape and present
enhanced states at the edges.  Figure~\ref{stm} $(a)$ shows the image
obtained at V$_s$ = 2~V. Here the distortion is evident for Mn$_5$ and Mn$_6$. Mapping
the conductance at a fixed bias of 1V gives direct evidence of the
localisation of the contributing electronic states. Indeed, Fig.~\ref{stm}
$(b)$ shows that most of the conductance is located on the borders of the
corresponding protrusions of Fig.~\ref{stm} $(a)$. However, no spatial
feature can be appreciated at negative bias.  Figure~\ref{stm} $(c)$
depicts the constant current image at -1V and a featureless protrusion
straddles the atoms of the chain. Consequently, the corresponding dI/dV
map (not shown) does not reveal any localisation inside the chain. 

In order to gain more insight, we plot in Fig.~\ref{stm} $(d)$ the conductance as a function of
bias for three different positions over the Mn$_6$ chain.  When the tip is
above a chain's edge, a distinct peak is detected
at 1V. This is in good correspondence with the previous Figs.~\ref{stm}
$(a)$ and $(b)$, and strongly suggests that there is an electronic state
localised at the edges of the Mn chains. When the bias is shifted to negative biases, there is a broader peak at $\sim -1$ V 
with larger intensity at the centre of the chain. From these data, we conclude that an electronic edge state appears at $\sim 1$ V,
while for occupied states an electronic state appears at $\sim -1$ V with an broader line shape, and extended along the chain.

%When the bias is shifted to negative biases, there is a peak at roughly
%$\sim -1$ V both when the tip is at the edge and at the centre of the
%chain, showing a small shift as the chain's edge is approached. For
%larger biases, all conductance traces show a behaviour akin to
%the free surface (also plotted) and we attribute them to the valence and
%conduction shoulders of the thin Cu$_2$N overlayer.

%From these data we conclude that an electronic edge state appears
%at $\sim 1 V$ and for occupied states an extended electronic
%state appears at $\sim -1V$ with some broadened features
%and shifting in energy due to the edge potentials of the Mn chains.

\begin{figure}[ht] 
\hspace{2cm}\hfill\includegraphics[width=0.9\columnwidth]{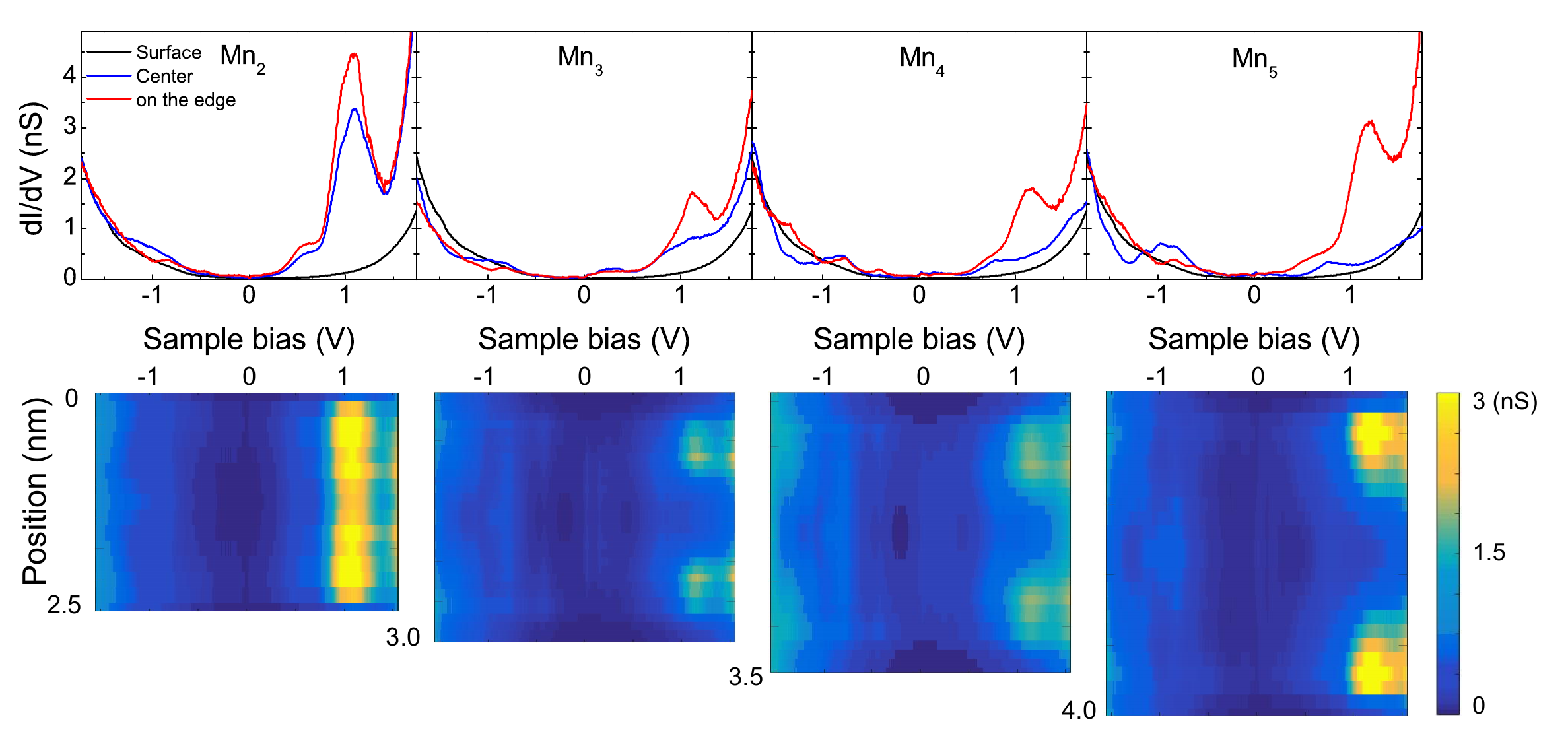}
\caption{\label{comp_didv} Tunnelling conductance over the edge (red),
and clean surface (black) for  Mn$_2$,  Mn$_3$,
 Mn$_4$, and  Mn$_5$, upper row (feed back opened at V$_s$ = -2~V, I$_t$ = 1.5~nA). Plots of
conductance (color scale) as a function of bias (x-axis) and position along
the chain (y-axis), the plots extend slightly more than the chain sizes.
All chains show the distinct feature of the edge states at the 
edges of each chain. }
\end{figure}

Figure~\ref{comp_didv} displays the tunnelling conductance as a function
of bias for Mn$_2$,  Mn$_3$, Mn$_4$, and Mn$_5$.  As the size of the Mn
chains is reduced, the occupied states evolve becoming very broadened and
undistinguishable from the conductance background, Fig.~\ref{comp_didv}.
On the contrary, the spectral intensity for the unoccupied state at the edges increases as the chain is reduced.
Moreover, the state stays at the same energy position, independently of the chain's length, supporting its localised character.

%Contrary to our expectation of an increasing interaction between edges
%as the chain is reduced in size, the edge state signal is fixed at $\sim$
%1 V, independent of the size of the chain, Fig.~\ref{comp_didv}.  However,
%the conductance through the edge state increases as the chain is reduced
%in clear contrast to the quenching of the occupied state.

\subsection{Density functional theory characterisation of the electronic states}
%Theoretical results

\begin{figure}[ht]
\centering
\includegraphics[width=0.9\columnwidth]{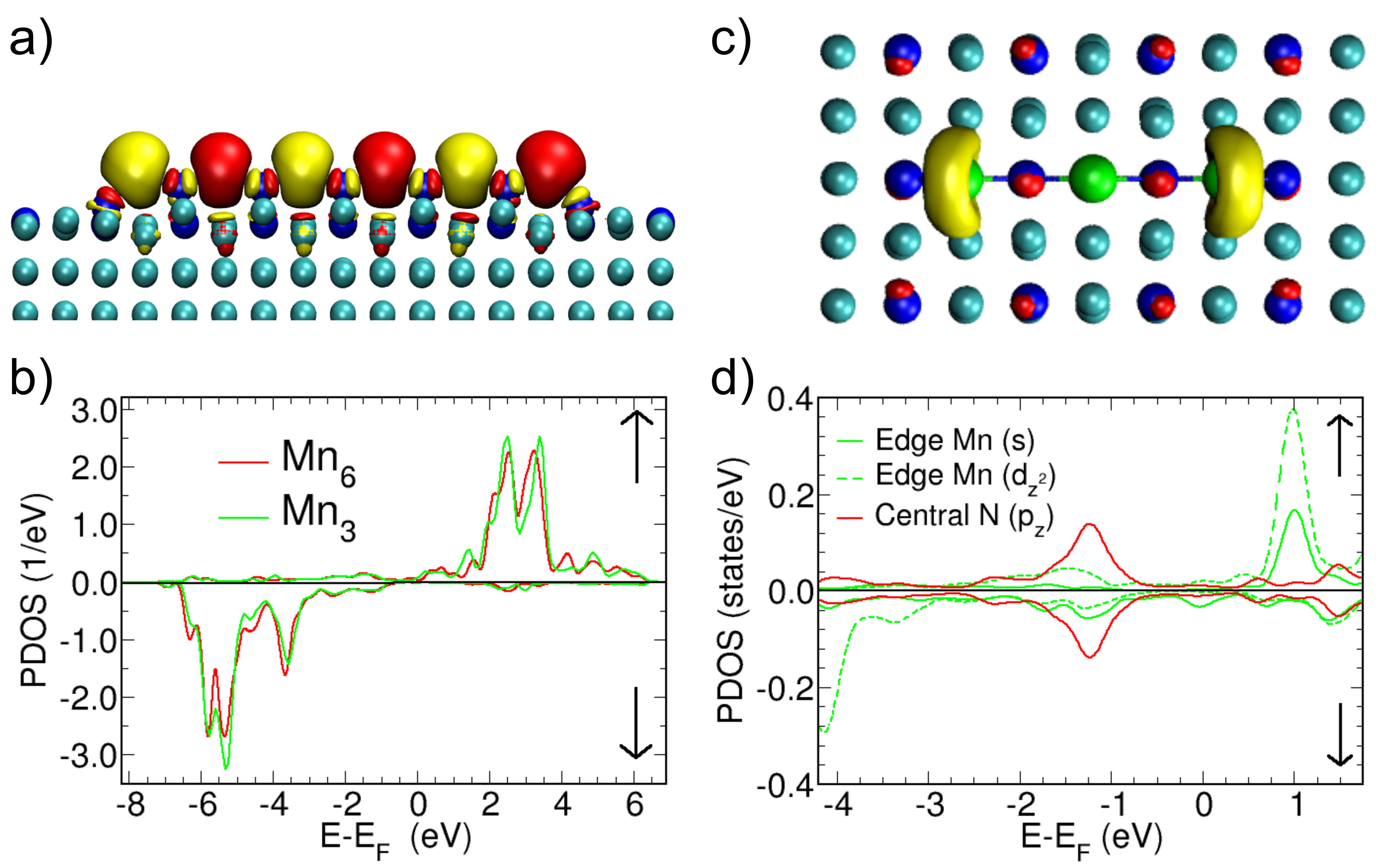}
\caption{\label{geom} 
$(a)$ Isosurface of spin density obtained as the difference
of electronic density between the densities of majoritary (red)
and minoritary (yellow) spins of a (broken-symmetry) DFT calculation.
$(b)$ Projected density of states (PDOS) over all Mn $d$-electrons of
Mn$_6$ and Mn$_3$ showing minor
differences, for the majority ($\downarrow$) and minority ($\uparrow$) spins.
$(c)$ Isosurface of wave-function amplitude of the edge state of an adsorbed Mn$_3$ chain.
$(d)$ PDOS of Mn$_6$ on the $p_z$ orbitals of N 
 (where $z$ is the direction along
the Mn$_6$ chain) and on the $s$ and $d_{z^2}$ electrons of a Mn edge atom. 
}
\end{figure}

Structural relaxation of the Mn atomic chains reproduce the geometry and bonding 
configuration from previous theoretical results~\cite{rudenko_2009,nicklas_2011,urdaniz_2012}:
Mn atoms induce an important reconstruction of the supporting substrate by 
incorporating N atoms to form a
Mn-N-Mn-N-~$\cdots$ chain. This has important consequences both for
the electronic structure and the magnetic ordering of the atoms. 
Figure~\ref{geom}$(a)$ shows isosurfaces of spin density.
 In agreement with
previous experimental studies~\cite{hirjibehedin_2006}, this corroborate
that  Mn atoms interact antiferromagnetically with their neighbours, 
as discussed by Rudenko
\textit{et al.}~\cite{rudenko_2009}, and also by Nicklas \textit{et
al.}~\cite{nicklas_2011} for the  case of Fe chains on the same substrate. The joining N atoms serve both
to stabilise the chain via covalent bonding with the Mn atoms, and to
induce the antiferromagnetic order through a superexchange interaction,
clearly seen by the equal coexistence of the two spins on the N atoms in 
Fig.~\ref{geom}$(a)$. These results imply that the actual arrangement
of Mn and N atoms does matter and different magnetic orderings can be
achieved~\cite{urdaniz_2012}.  These results also show that the occupied
electronic structure associated with N atoms must be spin unpolarised.

%we confirm that the magnetic ordering is antiferromagnetic, and as Rudenko
%\textit{et al.}~\cite{rudenko_2009} for Mn and Nicklas \textit{et
%al.}~\cite{nicklas_2011} for Fe discussed, the joining N atoms serve both
%to stabilise the chain via covalent bonding with the Mn atoms, and to
%induce the antiferromagnetic ordering through a superexchange interaction,
%clearly seen by the equal coexistence of the two spins on the N atoms,
%Fig.~\ref{geom}$(a)$. These results imply that the actual arrangement
%of Mn and N atoms does matter and different magnetic orderings can be
%achieved~\cite{urdaniz_2012}.  These results also show that the occupied
%electronic structure associated with N atoms must be spin unpolarised.

Figure~\ref{geom}$(b)$ plots the spin-polarised density
of states projected onto Mn $d$-states.  As previously
mentioned~\cite{rudenko_2009,lin_2011,nicklas_2011,urdaniz_2012}, the $d$
electrons maintain the free-atom configuration of Mn in the chains with
all majority spin $d$-orbitals occupied and the minority one empty.
Figure~\ref{geom}$(b)$ also shows that the PDOS on $d$-electrons is
mainly independent of the size of the Mn chains, indicating that
 their $d$-electron states are fairly localised and not perturbed by
neighbouring Mn atoms.

A consequence of the finite size of the chains is the apparition of
additional localised states at the terminations due to the  change of
geometry.  In strong correspondence with the experimental results, we
find  at V$_s \sim$1~eV above the Fermi energy a state purely localised
at the edges. This state,  depicted in Fig.~\ref{geom}$(c)$ for the case
of a Mn$_3$ trimer, is strictly spin-polarised, and has very little
weight on atoms other than the two edge Mn atoms. It thus has a small
intrinsic width.

The projected density of states gives us more information on the two states found in
the STM studies (Figs.~\ref{stm} and \ref{comp_didv}).  The edge states
only have contributions from  $s$ and $d_{z^2}$ orbitals (where $z$ is the
direction along the Mn chain). This leads to a sharp peak in the density
of states projected onto the $4s$ orbital of the edge Mn atom, centred at
$\sim 1$ V, Fig.~\ref{geom}$(d)$.  These data allow us to characterize
the edge state as a Tamm state due to the unsaturated $s-d{z^2}$ hybrid
orbital formed by the twisting of the chain at the edge.

The state at $\sim -1$ eV is also found in DFT if the full electronic
structure is projected onto the $p$ orbitals of the central N-atoms of
the chain. Figure~\ref{geom}$(d)$ shows a sharp peak in the PDOS of the
$p_z$ orbital of the third N atom in the Mn$_6$ chain. This allows us
to characterize the experimental peak at $\sim -1$ V as a chain state
originating in the N-atoms. The PDOS on the $p$ orbitals of the central
N-atoms is identical for both spins, as we expected for electronic states
with a strong N component. Hence, the experimental peak for occupied
states corresponds to a state extended over the chain with a strong
N character.

\begin{figure}[ht]
\centering
\includegraphics[width=0.7\columnwidth]{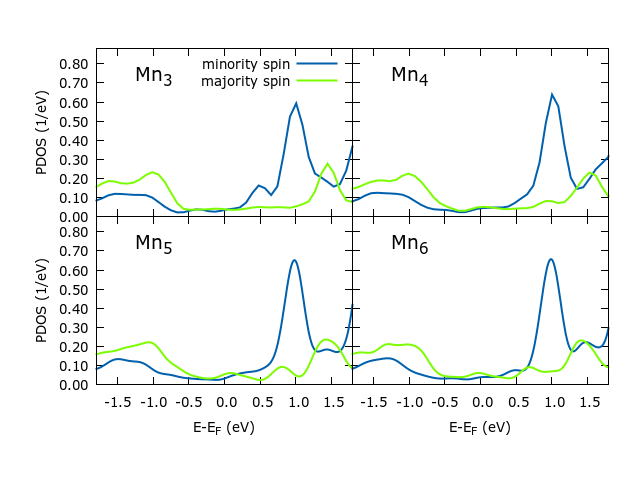}
\caption{\label{pdos} 
Projected density of states (PDOS) over all atomic orbitals of an edge
Mn atom for Mn$_3$, Mn$_4$, Mn$_5$, and Mn$_6$ for
majority and minority spins. The edge state
is pinned at the same energy and for the minority spin of each edge atom
in the DFT broken-symmetry picture. } 
\end{figure}

To explore the  evolution the edge states with chain length we compare
in Fig.~\ref{pdos}  the density of states  projected on an edge Mn atom
for Mn$_n$ chains with $n=3,4,5,6$.  In agreement with the experimental
results in Fig.~\ref{comp_didv}, the edge state is observed pinned at
$\sim 1$ eV and having basically the same shape regardless of the length
of the chain. This is due to the large localisation of the state at
the edge Mn atoms, thus interacting very weakly with the state at the
other end.  Even-numbered chains (Mn$_{2n}$ with $n$ integer) are an
interesting case because, in a broken-symmetry description, the two edge
states are of opposite spin and localised to each edge atom due to the
antiferromagnetic character of the chains.  The localisation of the edge
states due to their opposite magnetism holds even when preserving the full
entanglement of the antiferromagnetic solution. As a consequence, even for
the dimer, Mn$_2$, the two edge states are not interacting. However, in Fig.~\ref{comp_didv},
we observe that the conductance peak associated to the edge states increases for Mn$_2$.
For such a small chain, the STM tip can couple simultaneously to both edge states 
and, accordingly the conductance is expected to be larger.
%But this is only possible in an entangled spin structure of the chain: an electron can tunnel
%through both edges because both spins are present in the entangled solution at each edge.

%In the present case of a non-magnetic STM tip over a short chain, the conductance can
%nevertheless increase because the tip can couple simultaneously to both
%edges with each tunneling spin due to the reduced tip-edge of the chain distance.
%However, in the presence of spin polarisation, the conductance would
%only increase for smaller even-numbered chains because of the entangled
%structure of the spin of the chain. Indeed, an electron can tunnel through
%both edges because both spins are present in the entangled solution.

In Fig.~\ref{pdos}, we also observe that
the edge state spin
is anti-aligned with the spin of the edge atom (majority spin). However, previous
studies~\cite{Lorente2009,Novaes2010} showed that the electron
transmission proceeded through the majority spin due to the prevailance
of majority spin electrons at the Fermi energy.  Figure~\ref{pdos} shows
indeed that for all chains the majority spin density of states tends
to be larger, leading us to conclude that as for the single Mn atom,
electron transmission through the chains at low bias takes part mainly
in the majority-spin channel, but at large positive biases
the minority-spin components dominate the transmission. Let us notice that the electronic structure
of the single Mn atoms and chains of Mn atoms are subject to different
symmetry due to the clear axis of the Mn chains.

\section{Conclusions}

In summary, we have investigated the electronic structure of Mn atomic
chains contructed on Cu$_2$N/ Cu (100) by atomic manipulation.
We have found two electronic states in the tunnelling spectra:
an unoccupied Tamm state, very localised on the edge atoms and 
an occupied state extended along the chain. The unoccupied state 
presents a strict spin-polarisation, and is originated from the hybridization of Mn $d_{z^2}$ and $4s$ orbitals. 
The occupied state has weight on both N and Mn atoms and it
is not spin polarised due to the absence of magnetism of the N atoms.
We expect that in this model system, the parity of the number of atoms would have an effect in their spectral fingerprint.
For even-numbered Mn chains, their antiferromagnetic
character leads to a strict localisation of their edge states into a single
Mn edge atom because of the opposite spin-polarisation of the states
on each edge. This fact is independent of spin entanglement.  However,
for odd-numbered chains, there is no spin decoupling of the two edge states.
Nevertheless, the small interaction between neighbouring Mn atoms leads
to effectively decoupled edge states.

%When the chain contains an even number of Mn atoms, there is a 
%a strict localisation of the edge state to a single
%Mn edge atom because of the opposite spin-polarisation of the states
%on each edge due to the antiferromagnetic character of the chains. This fact is independent of spin entanglement.  However,
%for odd-numbered chains, there is no spin decoupling of the two edge states.
%Nevertheless, the small interaction between neighbouring Mn atoms leads
%to effectively decoupled edge states.

\section*{Acknowledgements}
DJC acknowledges the European Union for support under the H2020-MSCA-IF-2014 
Marie-Curie Individual Fellowship programme proposal number 654469. 
ICN2 acknowledges support from the Severo Ochoa Program (MINESCO, Grant SEV-2013-0295).

\hspace{1cm}
\bibliographystyle{unsrt}

\bibliography{mn}

\end{document}